# Zinc-Blende group III-V/group IV epitaxy: importance of the miscut

C. Cornet[1,*], S. Charbonnier[2], I. Lucci[1], L. Chen[1], A. Létoublon[1], A. Alvarez[1], K. Tavernier[1], T. Rohel[1], R. Bernard[1], J.-B. Rodriguez[4], L. Cerutti[4], E. Tournié[4], Y. Léger[1], G. Patriarche[5], L. Largeau[5], A. Ponchet[3], P. Turban[2] and N. Bertru[1]

*[1]Univ Rennes, INSA Rennes, CNRS, Institut FOTON – UMR 6082, F-35000 Rennes, France*
*[2]Univ Rennes, CNRS, IPR (Institut de Physique de Rennes) - UMR 6251, F-35000 Rennes, France*
*[3]CEMES-CNRS, Université de Toulouse, UPS, 29 rue Jeanne Marvig, BP 94347 Toulouse Cedex 04, France*
*[4]IES, Univ. Montpellier, CNRS, Montpellier, France*
*[5]Centre de nanosciences et de Nanotechnologies, site de Marcoussis, CNRS, Université Paris Sud, Université Paris Saclay, route de Nozay, 91460 Marcoussis, France*

Here, we clarify the central role of the miscut during group III-V/ group IV crystal growth. We show that the miscut first impacts the initial antiphase domain distribution, with two distinct nucleation-driven and terraces-driven regimes. It is then inferred how the antiphase domain distribution mean phase and mean lateral length are affected by the miscut. An experimental confirmation is given through the comparison of antiphase domain distributions in GaP and GaSb/AlSb samples grown on nominal and vicinal Si substrates. The antiphase domain burying step of GaP/Si samples is then observed at the atomic scale by scanning tunneling microscopy. The steps arising from the miscut allow growth rate imbalance between the two phases of the crystal and the growth conditions can deeply modify the imbalance coefficient, as illustrated with GaAs/Si. We finally explain how a monodomain III-V semiconductor configuration can be achieved even on low miscut substrates.

The monolithic integration of both Zinc-Blende and Wurtzite III-V semi-conductors respectively on (001) and (111) group-IV substrates (such as Si or Ge), is nowadays one of the most promising approach for the development of integrated photonic devices, or efficient energy production and storage applications [1–3]. More specifically, (001) substrates are generally preferred over (111) ones, as it is expected to ease the post-growth processing of group III-V/ group IV devices [1]. On the other hand, crystal defects generated in III-V epilayers grown on group IV substrates may be numerous and detrimental for devices operation. Especially, antiphase domains (APDs) which are related to the polar on non-polar epitaxy (i.e. to the two different ways for the III-V crystal to fit the group IV substrate orientation) are strongly impacting the structural, and electronic properties of grown III-V semiconductors. The easiest solution to avoid or mitigate the formation and propagation of antiphase boundaries (APBs) through devices is to grow the III-V materials on misoriented group IV (001) substrates. But the post-growth processing of such misoriented (vicinal) III-V/IV wafers remains tricky [1], especially when the miscut angle reaches 1° or more. Recently, many research groups tried to reduce or even suppress the miscut of the used group-IV wafer. [4] But a clear view on the relationship between miscut, APBs generation, and APBs propagation is still missing.

Indeed, the use of a vicinal substrate is often motivated by the ability to promote the double step formation at the group IV surface [5], avoiding the monoatomic layer translation of the III-V crystal that may appear due to the presence of single steps at the substrate surface, theoretically generating an APB. With this picture in mind, K. Volz *et al.* explained their results about epitaxial GaP/Si by considering a 2D III-V growth mode on the substrate [6] (Note that the difference between a 2D growth mode and a flat 3D one is difficult to make experimentally).

On the other hand, the recent work of I. Lucci *et al.* proposes an alternative model [7], by demonstrating the following points from Density Functional Theory and extensive experimental observations on GaP/Si, AlSb/Si and AlN/Si: (i) There is only a partial wetting between III-V semiconductors and Si, thus leading to the formation of pure 3D Volmer-Weber growth mode, as confirmed experimentally by some other studies [8,9]. AlSb/Si islands have even been found to be at their equilibrium shape. [8] Given the surface energy orders of magnitude, and the results of the literature [10], this also applies to III-V/Ge. (ii) APBs are generated during the heterophase coalescence of 3D islands, as also suggested in pioneering works for III-V/Ge [10]. The size of individual monodomain islands can be much larger than the distance between steps (the terrace width) [8]. The steps were found to have no impact on APDs generation. (iii) Elastic energy does not have a significant impact on the island morphology at the coalescence growth step (and therefore does not impact the APD distribution), as most systems are already plastically relaxed (even for GaAs/Si [11]) or quasi-lattice-matched.

(iv) The epitaxial relationship is defined locally at the nucleation site that further governs the phase distribution.

Meanwhile, extensive works were performed about the dislocation-free GaP/Si model case. A clear correlation between the Si local surface dimer orientation and the subsequent epitaxial III-V phase was established (See for instance ref. [12] or [13] and references therein). This led the authors to support the idea of a step-induced generation of APBs. Although these observations may seem in contradiction with the 3D nucleation described previously, the connection will be established later on in this work. Finally, many groups tried to favor the so-called APB annihilation by playing with III-V growth parameters. It was noticed that the V/III ratio and growth temperature play an important role in this process [14,15], suggesting a significant contribution of kinetic effects at this step.

In this work, we aim to clarify the impact of the miscut on the generation of antiphase domains during the heteroepitaxy of group III-V semiconductors on group IV substrates. We first investigate the influence of the miscut on the initial APD distribution, in the low or large miscut regimes. The APD burying is then considered from the point of view of the growth rate imbalance of the different III-V phases. The ability to bury APD with low miscut substrates is finally discussed.

In order to clarify the impact of the miscut on the initial distribution of III-V islands phase we first review the results of the literature. For low miscut Si substrates (typically <1°), Beyer et al. [13] have demonstrated that the APD distribution in the GaP crystal reproduces well the distribution of steps at the Si surface. For large miscut (typically >1°), on the contrary, many studies report on APDs size significantly larger than average terraces width. For the MOCVD GaP growth on 2°-off Si (i.e. 3.89 nm expected average terrace width), 9 and 26 nm-large APDs were observed [16]. For the MBE GaP growth on Si-4°-off (i.e. 1.94 nm expected average terrace width), a complete analysis of APD distribution lead to the conclusions that APDs lateral width lye between 10 and 58 nm [15]. Finally, on Si-6°-off substrates (i.e. 1.29 nm expected average terrace width), 12 nm-average size APDs were found for the MBE growth of GaP [7,17,18]. From all these results, it therefore appears that while the APDs are related in some extent to group IV steps when the III-V growth is performed on low miscut substrates, it is absolutely not the case when the III-V growth is performed on large miscut group IV substrates.

Also, it was demonstrated that partial wetting conditions are theoretically expected and experimentally observed for group III-V/group IV heterogeneous growth [7,8]. Therefore, it is reasonable to postulate that the III-V growth on group IV substrate starts with the formation of monodomain III-V 3D islands, which phase is defined directly by the surface dimer orientation of the substrate terrace where the III-V island nucleation occurs. With this hypothesis, it is easily understandable that for low miscut group IV substrates (see Fig. 1(a)), terraces are large, and therefore a number of III-V islands with the same phase will grow on the same terrace. After coalescence of these islands, this will lead to a "terraces-driven" APD distribution, where the APBs will follow approximately the monoatomic step distribution at the group IV substrate surface. But it is only an approximation, as before coalescence individual islands may grow over a Si monoatomic step while staying monodomain, as schematically represented in Fig. 1(a). Thus, for low miscut substrates, the lateral size of APDs is expected to be directly related to the average lateral size of the substrate surface terraces. For large miscut group IV substrates (see Fig. 1(b)), terraces are small, and the size of each individual monodomain III-V island is larger than the width of terraces. For illustration, on Si-6°-off substrates (i.e. 1.29 nm expected average terrace width), 10-50 nm large islands were determined for the MBE growth of AlSb/GaSb [8]. This will lead to a "nucleation-driven" APD distribution, where the size of APDs after coalescence is mainly related to the distance between two neighboring islands of opposite phases generated during nucleation.

The frontier between the terraces-driven and the nucleation-driven APD generation regimes corresponds to the situation where the average width of the terraces at the group IV substrate surface equals the average distance between two neighboring islands of opposite phases along the miscut direction. For given growth conditions, a given miscut allows reaching this situation, named hereafter the "critical miscut".

The average distance between two neighboring islands along the miscut direction can be extracted from the island surface density d measured experimentally by considering a Poisson distribution of the III-V islands positions at the group IV surface [19] :

$$d_{1D} = \frac{1}{\pi \sqrt{d}} \qquad (1)$$

Figure 1(c) represents the evolution of the average terrace width as a function of the miscut angle, for a silicon surface composed of monoatomic steps. We note here that this trend is also valid for a Ge surface. At the scale of the process described here, the difference is so small that the evolution for Ge can be considered as similar to the one of Si. For illustration purposes, average distance between two neighboring islands along the miscut direction were extracted for different III-V/IV systems studied in the literature: GaSb/Si [20], AlSb/Si [20], GaP/Si [7], InP/Si [21], GaAs/Si [22,23] and GaAs/Ge [10]. These values are superimposed with the terraces width curve in Fig. 1(c). In these previous works, islands are observed well after the nucleation step. Therefore, the island density is not expected to change during the growth, and would remain the same for thicker layers at the coalescence step.

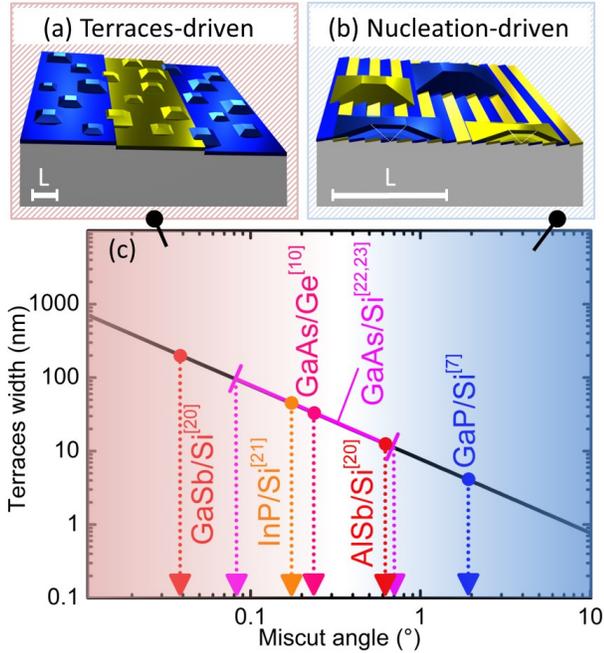

FIG. 1: Sketch of (a) the terraces-driven phase distribution in III-V islands grown on low miscut group IV substrates, and of (b) the nucleation-driven phase distribution in III-V islands grown on large miscut group IV substrates. Blue and yellow colors are used to indicate the different surface dimer orientations and phases of both group IV terraces and III-V islands. L indicates the mean size of III-V islands. (c) Average length of terraces along the [110] or [1-10] directions as a function of the miscut angle for Si or Ge surfaces composed of monoatomic steps. Colored dots correspond to the 1D average distance between islands reported for different III-V/group IV systems, from refs. [7,10,20–23]. The vertical dashed arrows indicate the corresponding critical miscuts.

On the other hand, the initial density of stable nucleii depends on the growth conditions used. Fig. 1(c) thus gives the corresponding critical miscuts between the two APD generation regimes for each materials system in given growth conditions. Red and blue miscut ranges are highlighted in Fig. 1(c) and are regions where terrasses-driven and nucleation-driven APD distribution are likely to occur. Of course, number of parameters could have an impact on the critical miscut, including growth conditions used during the nucleation, but also the real number of steps at the surface and the growth technique chosen (e. g. Molecular Beam Epitaxy (MBE) or Metal-Organic Chemical Vapor Deposition (MOCVD)). Nevertheless, most of the previous works report island surface densities in the $[10^9-10^{11}]$ cm$^{-2}$ range, that allows to conclude that the critical miscut is in the [0.1-1] ° range for common group III-V/group IV heterogeneous associations. But lower values for the critical miscut can be achieved by changing the nucleation conditions, and thus the initial island density [24].

Overall, the initial APD distribution (very near the III-V/group IV interface) can be fully characterized by two parameters. First, the III-V crystal mean phase should be considered, ranging from -1 to +1 [17,25]. Here, for APBs propagating vertically, a mean phase of 0 means equal number of atoms in the main phase and antiphase domains. The island nucleation being a stochastic process, the mean phase is directly related to the area ratio between the two different group IV substrate terraces local surface dimer orientations. Note that for a given III-V crystal mean phase, many different monoatomic or biatomic group IV steps possible configurations may be considered. Inversely, a given density of monoatomic steps at the substrate surface is not enough to predict the mean phase of the III-V crystal. Therefore, the achievement of statistically-dominant biatomic steps distribution at the surface certainly helps to promote a near to single phase domain configuration (i.e. a mean phase of +1 or -1) in III-V layers [26].

The second important parameter is the mean lateral extent of APDs, which is related either to the terrace width below the critical miscut, or to the nucleation islands density above the critical miscut. Here, it should be mentioned that the concept of critical miscut was introduced by considering a perfect monoatomic step lattice at the group IV surface. But, depending on the strategy used to prepare the group IV substrate (e.g. chemical preparation or homoepitaxy), the real step distribution can be quite different than the ideal one. Especially, for low miscut substrates, a fine control of the miscut angle and the miscut direction in addition to a proper homoepitaxial or passivation strategy is needed to reach the perfect terraces-driven APD distribution regime over the whole sample [13].

In Fig. 2, a comparison between similar thick III-V layers grown on low miscut and large miscut group IV substrates is shown. Growth and microscopy details are given in the supplemental materials [19]. At first, cross-sectional Transmission Electron Microscopy (TEM) images of two comparable samples mainly composed of GaSb (with a thin AlSb nucleation layer) grown on freshly prepared Si 0.3°-off along the [110] direction (Fig. 2(a)) and 6°-off along the [110] direction (Fig. 2(b)) are shown. In these images, bright and dark contrasts mainly correspond to the main phase or antiphase domains contrasts, although the presence of other defects may contribute as well. At first sight, it can be seen that APDs are overall larger and higher for the low miscut case than for the large miscut case.

More specifically, the images show two distinct features: (i) the presence of very small APDs very near the interface, with a width typically lower than 20nm that correspond to individual islands formed initially [7]. (ii) Some large APDs propagating over 50-100 nm thicknesses and beyond, that corresponds to the APD distribution after the burying of some of the previously discussed small APDs. This apparent bimodal distribution may be related to the adatom diffusion length needed for APD burying, that will be discussed later in this work.

Considering the initial APD distribution, *i.e.* near the III-V/Si interface, it can be seen that while the GaSb growth on 6°-off leads to very small dark and bright area observed all along the interface, the growth on low miscut substrate gives rise to large single-phase areas between two APBs (up to 40 nm for the low miscut case vs up to 15 nm for the large miscut case). This observation is in good agreement with the previously proposed explanation of the impact of the miscut on initial APD distribution. Now looking at large APDs, it can be seen that they are smaller on 6°-off substrate than on the (001) one. Furthermore, a monodomain GaSb layer is finally reached for the vicinal case, while APBs are propagating until the surface for the nominal one.

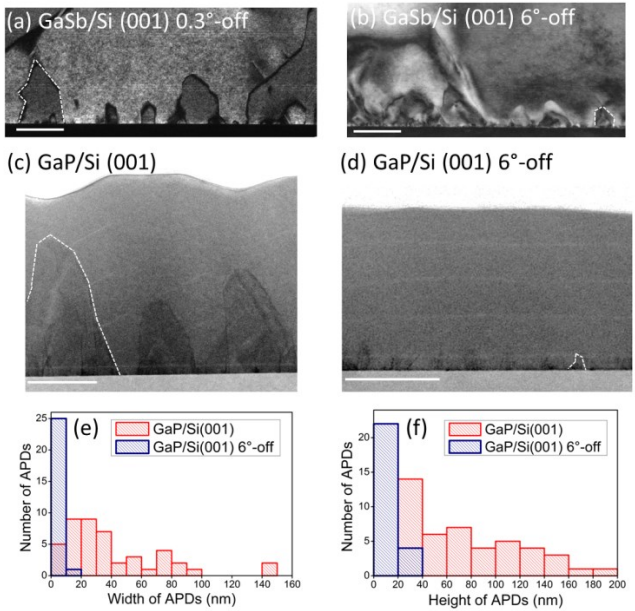

FIG. 2: Cross-sectional Transmission Electron Microscopy images of thick (a) GaSb/Si(001) 0.3°-off, (b) GaSb/Si(001) 6°-off, (c) GaP/Si(001) and (d) GaP/Si(001) 6°-off samples along the [110] direction. The white bar represents 100nm. Dashed lines are guide to the eyes showing typical APBs shape in the sample. Corresponding antiphase domains width (e) and height (f) distributions are quantitatively represented for GaP/Si samples on both nominal and vicinal substrates.

But a meaningful quantitative assessment cannot be given with these pictures, because of the interplay between antiphase boundaries and other defects, such as dislocations [27]. A more ideal case can be found with the quasi-lattice-matched GaP/Si system. Cross-sectional Transmission Electron Microscopy (TEM) images of two comparable samples (with the same growth conditions) are mainly composed of GaP grown on freshly prepared Si (001) (0 +/- 0.5°) substrate (Fig. 2(c)) and 6°-off along the [110] direction (Fig. 2(d)) are shown. It is useful to recall that the exactly oriented nominal group IV (001) substrate is a theoretical case, not achievable by substrate manufacturers. In practical cases, a residual and often uncontrolled miscut should always been considered [6]. Bright and dark contrasts are again attributed to domains with different phases. A statistical analysis was performed over a cumulated width of 4 μm along the [110] direction (see examples of TEM images in the supplemental materials [19]). Results obtained for the APDs width and heights are given in Fig. 2 (e) and (f). Emerging APBs observed for the (001) case are not taken into account for this statistics. A perfect deconvolution between small APDs and large APDs populations is not achievable, but the images and statistics given clearly show that small APDs, near the III-V/Si interface are more numerous and smaller on 6°-off substrates, that confirms the impact of the Si terraces width in the low miscut limit. We also confirm quantitatively that large APDs are much larger on the nominal substrate, and propagate farther in the sample. Finally, Fig. 2(c) also points out that APBs induce roughness and faceting [28] with consequences well after their annihilation. Indeed, it can be seen on the 3 APBs shown in Fig. 2(c) that the flatness of the free surface is directly related to the distance to the highest point of the buried APD. After a given thickness, the (001) surface is recovered, as shown for the central APD of Fig. 2(c). That is why a thin buffer layer is then needed after APBs annihilation to smoothen the surface. Overall, these images confirm the central role of the miscut on the initial APD distribution, but also point out the importance of the miscut for subsequent steps of III-V/Si growth.

Therefore, the role of the miscut in the so-called APB annihilation process needs to be clarified. Numbers of situations were reported in the literature ranging from annihilation achieved on large miscut group IV substrates [25] or nominal substrates with a residual miscut (<0.2°) [6], by MBE [15], or by MOCVD [16]. In these references, and in the Fig. 2 TEM images, various APBs profiles are observed, with single or many facets composing the APB. A more precise idea of the mechanisms involved during the "annihilation process" can be obtained by imaging the morphology of the surface precisely at the moment where one domain becomes statistically dominant over the other one. To this aim, a 200-nm thick GaP layer was grown on a Si (001) − 6°-off substrate (see the supplemental materials [19] for growth details) in growth conditions, different from those used in samples shown in Fig. 2(d), such that GaP would become purely monodomain at around 300 nm. The sample was then transferred to the Scanning Tunneling Microscope (STM) chamber [19,29] for further surface investigation at the atomic scale. The STM image obtained is shown in Fig. 3(a). The local crystallographic directions of the two different GaP phases can be distinguished at the atomic scale by the surface reconstructions, as shown in Fig. 3(a) inset. At a microscopic scale, the two phases are also distinguishable as the surface is composed of elongated domains either along the [110] direction of the Si substrate, or along its [1-10] direction. Before coalescence a domain is always elongated along its own [-110] GaP local direction. After the coalescence, the statistically-dominant domain is elongated along its own [-110] GaP local direction.

But the most interesting feature observed in Fig. 3(a) is the way one domain dominates the other one. Especially, it can be seen that the GaP domains having their [-110] direction parallel to the [110] direction of the Si substrate seem to coalesce over the other domains, through the development of thin "bridges", leading to continuous and elongated single phase domains at the surface. We note here that for similar growth conditions, and similar substrates, the dominant phase is always the same, on different part of the samples, and even for different samples, as shown in the supplemental materials [19].

From this picture, it is now clear that annihilation of APBs is simply the result of the antiphase domains burying. But this view also implies that the two different domains composed of the same material, have different growth rates. This is somehow surprising as the two domains present the same (001) surfaces at the growth front. This is where the substrate miscut plays an important role. Indeed, it breaks the symmetry between the two different III-V domains. Fig. 3(b) illustrates the impact of a miscut along the [110] direction for a group IV substrate on the III-V layers grown on it. For the sake of simplicity, a (2x4) surface reconstruction is schematically represented at the top surface of the III-V semiconductor. Atomic reconstruction of steps is not represented here. Especially, it can be seen that for one III-V domain, the miscut is transferred along the [110] direction, while it is transferred to the [-110] direction for the other domain. Consequently, if the surface of one domain has more A-steps (having its edge parallel to the group V dimers, illustrated by the yellow color in Fig. 3(b)), the surface of the other domain will have more B-steps (having its edge perpendicular to the group V dimers, illustrated by the blue color in Fig. 3(b)) [30]. The incorporation rate on A- and B-steps being different [31,32], this is how growth rates of domains having different phases will be different. Here it is important to mention that the illustration only considers the case of vicinal (001) III-V surfaces. High angle facets formed at the surface where the APB emerges (see the supplemental materials [19]) could be as well represented. But depending on the different materials systems and miscuts, large other stable facets (see for instance the presence of (114) ones for GaP/Si in ref. [28]) may develop, especially at the APD edges. The presence of these facets will certainly have an impact on the growth rates of each domain.

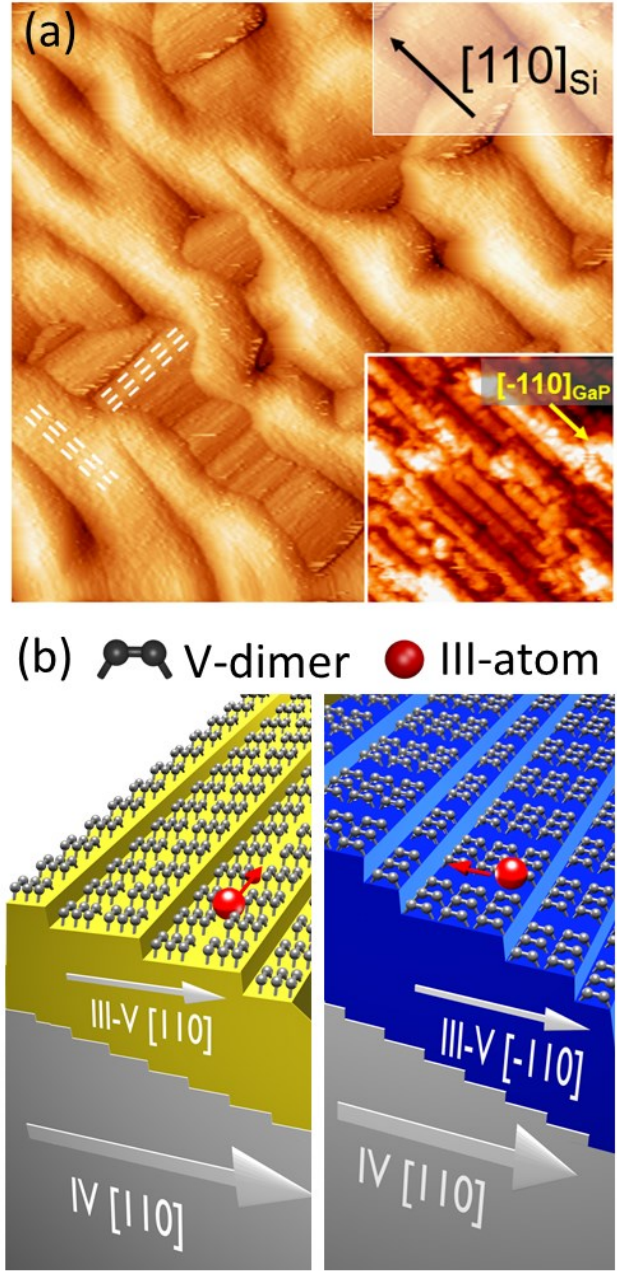

FIG. 3: (a) Plan-view STM image of a 200 nm-thick GaP deposition on Si (001) – 6°-off (400*400 nm2, vertical color scale: 0-13.9nm), during the APB annihilation step. The 19*19 nm² inset shows the atomically-resolved morphology of the GaP dominant phase. Dashed lines are guide to the eyes showing local different GaP phases. (b) Illustration of the asymmetry induced by the miscut on the different III-V phases grown on a group IV substrate, enabling different growth rates for the different III-V phases.

In the following, we will call α and β the two different phases of the III-V crystal having respectively more A-steps and B-steps at the surface. The phase α results from the III-V nucleation on a $T_\alpha$ terrace at the Si surface, while the phase β results from the III-V nucleation on a $T_\beta$ terrace at the Si surface. $T_\alpha$ and $T_\beta$ are Si terraces with different surface dimers orientations. The determination of the growth rate imbalance between the α- and the β- phases is therefore of interest, as the early burying of APDs is requested for highly integrated photonics [1] while the propagation of APBs is of interest for some non-linear photonic [25] or water splitting applications [3]. The growth of III-V semiconductors on miscut surfaces may follow two different growth modes, namely step-flow growth or 2D-nucleation ones, or a combination of both [33]. Especially, for given growth conditions, with a very low miscut, the distance between steps at the surface is so large that the growth depends mostly on the diffusion length of adatoms at the surface and 2D-nucleation growth mode arises. For larger miscuts, the incorporation of group III atoms at the step edges becomes dominant over all the other contributions, and the step-flow regime is reached. This regime corresponds to the situation where the distance between steps is lower than the diffusion length of group III adatoms at the growth temperature considered. [33] The typical values of the diffusion lengths given for group III adatoms on a III-V planar surface are in the [0.5-1] µm range [14], implying that the miscut at which step flow growth mode may occur is typically larger than [0.01-0.03]°, depending on the material system and growth conditions. In the present case, a pure 2D nucleation growth mode would give the same growth rates for the α- and β- phases. Some fraction of step-flow is therefore required to break the growth rate balance between α- and β- phases. Therefore, on can consider that burying of APDs is only possible if the substrate has at least a projected miscut of 0.03° precisely along a given [110] or [1-10] direction. Beyond this value, the respective growth rates of phases α and β only depend on the incorporation rates at steps A and B.

We propose to introduce the growth rate imbalance coefficient $C_{\alpha/\beta}$ defined as:

$$C_{\alpha/\beta} = \frac{V_{g\alpha}}{V_{g\beta}} = \frac{R_A^0 \cdot e^{-\frac{E_A}{k_B T}}}{R_B^0 \cdot e^{-\frac{E_B}{k_B T}}} \qquad (1)$$

Where $V_{g\alpha}$ and $V_{g\beta}$ are the crystal growth rates of the phases α and β. $R°_A$, $R°_B$, $E_A$ and $E_B$ are respectively the amplitudes and energy barriers for direct incorporation at steps A and B, as defined in refs. [31,32]. $k_B$ is the Boltzmann constant, and T the growth temperature. Therefore, the growth rate imbalance coefficient is a simple ratio between the growth rate of phases α and β, which is equal to the ratio of direct step incorporation rate per site for each phase. Therefore, if $C_{\alpha/\beta}$ is lower than unity, the β phase will grow faster than the α one. On the contrary, if $C_{\alpha/\beta}$ is larger than 1, The α phase will become dominant. If

$C_{\alpha/\beta}$ equals to one, the growth rates will remain equal and APBs will propagate to the surface. Interestingly, the development of the α- or β- phases and the rate at which it occurs does not depend on the miscut angle. Indeed, a given miscut angle will define the same areal density of steps at the surfaces of α- or β- phases. APD burying is therefore expected to occur in the same way on low and large miscut group IV substrates.

The determination of the growth rate imbalance coefficient is however conditional upon knowing experimentally the growth rates or direct step incorporation rates per site for each phase α and β. Experimental determination of incorporation rates was proposed in the pioneering works of Shitara et al. for MBE-grown GaAs, by using reflection high-energy electron diffraction [31,32]. From these data (see supplemental materials for the parameters used [19]), the growth rate imbalance coefficient $C_{\alpha/\beta}$ was calculated and plotted in Fig. 4 as a function of the temperature, for various V/III ratio. Results shown in Fig. 4 can be applied directly to the MBE growth of GaAs on Si or Ge substrates and in some extent transposed to MOCVD GaAs/Si or GaAs/Ge growth.

First, Fig. 4 shows that for most growth conditions, the β phase will grow faster than the α one, except at high growth temperature and low V/III ratio, where the α phase will be favored. It is also noticed that the lower the V/III ratio, the more the $C_{\alpha/\beta}$ coefficient becomes sensitive to the growth temperature. As a consequence, the two extreme values of $C_{\alpha/\beta}$ leading to the most important differences between growth rates are achieved at high and low temperatures respectively, but always at a low V/III ratio. This explains the experimental observations given by I. Lucci et al. [28]. In this previous work, thick GaP/Si samples were grown, resulting in a single domain GaP in the central part of the wafer, while at the edges of the 2" wafer, another GaP single domain with opposite phase was observed. The local increasing of the temperature at the edges of the wafer has certainly allowed changing the dominant phase.

It is also interesting to see that for a given growth temperature (high enough), the V/III ratio may allow to tune the dominant phase at will. This is illustrated in Fig. 4 inset, where the $C_{\alpha/\beta}$ coefficient was plotted as a function of the V/III ratio for GaAs at a growth temperature of 620°C. With increasing V/III ratio, the dominant phase changes from α to β. Even if the trends described for GaAs in Fig. 4 are expected to be similar for other III-V semiconductors and thus provide a guide for the growers, a precise determination of direct step incorporation rates for the different materials systems are still needed to optimize precisely the heterogeneous group III-V on group IV epitaxy on purpose, with each material specificity and surface or step reconstructions. The hydrogenated surface during MOCVD III-V growth may also be able to impact significantly the step incorporation rates imbalance.

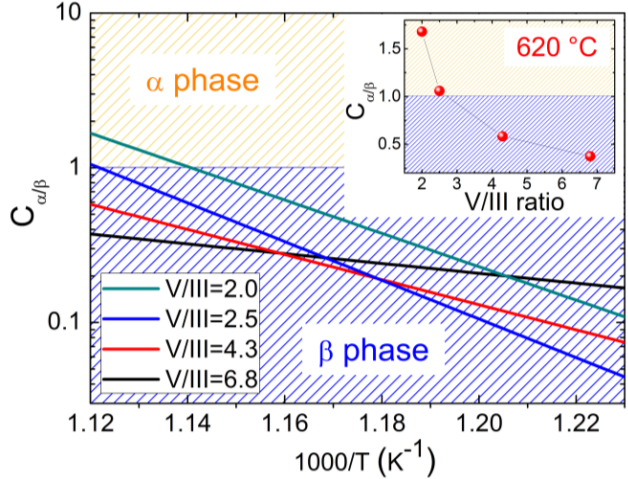

FIG. 4: Growth rate imbalance coefficient $C_{\alpha/\beta}$ as a function of 1000/T for various V/III ratio used in the case of GaAs/group IV epitaxy, determined from ref. [31,32]. The orange area indicates the conditions where the α III-V phase grows faster, while the blue area indicates the conditions where the β III-V phase grows faster. Inset shows the evolution of the imbalance coefficient as a function of the V/III ratio for GaAs grown at 620°C on a group IV substrate.

The consequences of this antiphase domain burying process description are numerous but we would like to highlight the most important ones. First of all, the mean phase of the initial APD distribution does not impact on the final phase of the layer. Even if the initial $T_\alpha/T_\beta$ surface ratio is at 80/20, subsequent growth conditions favoring the β phase will lead to a monodomain β-phase crystal after a sufficient thickness. Second, the thickness at which a single domain III-V semiconductor is recovered is mainly dependent of the initial APD mean lateral extent, and the growth rate imbalance coefficient $C_{\alpha/\beta}$. A large initial APD, as the one observed on nominal substrates in Fig. 2(a) and (c) or in ref. [6], will require a larger deposition thickness to be buried (see fig. 2 e) and f)). In the same way, large deposition thicknesses will be needed before the APD burying if growth conditions corresponding to a near to 1 $C_{\alpha/\beta}$ coefficient are chosen. Third, after the initial APD distribution generation, the general morphology (and "facets") of APDs is governed by the growth rate imbalance coefficient $C_{\alpha/\beta}$. APBs lying along crystallographic directions far from the [001] one typically traduce a strong growth rate imbalance, i.e. $C_{A/B}$ much larger or lower than 1. On the contrary, vertical propagation of APBs indicates a $C_{\alpha/\beta}$ coefficient close to 1 or a too low miscut angle, at least locally. Of course, this general picture does not allow to predict the APB structure at the atomic level, that may be locally impacted by charge compensation effects or temperature-induced kinks, as described by Beyer et al. [34]. Last, from this description it can be understood why burying of APDs can be achieved even on low, but controlled, miscut substrates, if the growth conditions, and especially V/III ratio and growth temperature are carefully chosen to promote growth rate imbalance. Here, we point out that a successful APD burying achieved on a low miscut group IV substrate requires the precise control of both miscut angle (larger than 0.03° to keep the step flow growth mode) and miscut direction, homogeneously at the substrate surface, so that the growth rate imbalance is achieved everywhere in the sample, despite some local miscut direction or angle fluctuations.

In conclusion, the link between the group IV substrate miscut, and the initial III-V antiphase domain distribution mean phase, and mean lateral size during III-V/group IV epitaxy was clarified. The central role of the miscut in the antiphase domain burying was established by taking into account the growth rate imbalance between the two III-V crystal phases. On this basis it was shown how burying of antiphase domains is possible for low miscut substrates. The detailed description of the group IV substrate miscut impact on epitaxially grown III-V structural properties opens new prospects for the development of highly integrated photonics or energy production/storage applications.

The authors acknowledge Dr. R. Leguevel and A. De Verneuil for fruitful discussions on the statistical analysis of experimental data for the determination of the critical miscut. Pr. K. Volz or Dr. J.-C. Harmand are also acknowledged for the interesting discussions about III-V/ group IV epitaxial processes. The authors acknowledge RENATECH (French Network of Major Technology Centers) within Nanorennes for technological support. This research was supported by the French National Research Agency ANTIPODE Project (Grant no. 14-CE26-0014-01), ORPHEUS Project (Grant no. ANR-17-CE24-0019-01) and Région Bretagne.

# Zinc-Blende group III-V/ group IV epitaxy: importance of the miscut Supplemental Materials

C. Cornet[1,*], S. Charbonnier[2], I. Lucci[1], L. Chen[1], A. Létoublon[1], A. Alvarez[1], K. Tavernier[1], T. Rohel[1], R. Bernard[1], J.-B. Rodriguez[4], L. Cerutti[4], E. Tournié[4], Y. Léger[1], G. Patriarche[5], L. Largeau[5], A. Ponchet[3], P. Turban[2] and N. Bertru[1]

[1]Univ Rennes, INSA Rennes, CNRS, Institut FOTON – UMR 6082, F-35000 Rennes, France

[2]Univ Rennes, CNRS, IPR (Institut de Physique de Rennes) - UMR 6251, F-35000 Rennes, France

[3]CEMES-CNRS, Université de Toulouse, UPS, 29 rue Jeanne Marvig, BP 94347 Toulouse Cedex 04, France

[4]IES, Univ. Montpellier, CNRS, Montpellier, France

[5]Centre de nanosciences et de Nanotechnologies, site de Marcoussis, CNRS, Université Paris Sud, Université Paris Saclay, route de Nozay, 91460 Marcoussis, France

## DETERMINATION OF THE CRITICAL MISCUT

At the beginning of III-V/Si crystal growth, 3D III-V islands are formed at the Si surface. The surface density of these islands is usually directly inferred from direct Atomic Force Microscopy or Transmission Electron Microscopy techniques. From this value, one can directly determine the average distance between two islands. Here, the situation is different, as determining the critical miscut requires the knowledge of the average distance between islands in a specific crystallographic direction (e.g. the [110] one), corresponding to the miscut direction. In the following, the formation of stable III-V nuclei at the Si surface are considered as independent and random in space (as a first approximation). A description of III-V islands surface distribution can thus be given by using the Poisson distribution. Therefore, considering d to be the surface density of islands, the probability to be at a distance r of an island in $\mathbb{R}^2$ is:

$$P(r) = 2\pi r d e^{-d\pi r^2} \quad (1)$$

The mathematical expectation in $\mathbb{R}^2$ is obtained by averaging this value over r, and gives the mean distance between two islands:

$$d_{moy} = \int_0^\infty P(r) r dr = 2\pi d \int_0^\infty e^{-d\pi r^2} r^2 dr \quad (2)$$

By integration, it comes:

$$d_{moy} = \frac{1}{2\sqrt{d}} \quad (3)$$

If we consider two islands aligned toward a direction having an angle θ with the miscut direction, the average distance along the miscut direction is therefore:

$$d_{1D,\theta} = d_{moy}|\cos\theta| \quad (4)$$

Finally, assuming a perfectly isotropic in-plane island distribution (corresponding to experimental observations), one can average this value over all the θ angles between 0 and 2π, in order to get the mean distance between 2 islands in 1D :

$$d_{1D} = \frac{1}{2\pi} \int_0^\infty d_{moy}|\cos\theta| d\theta = \frac{1}{\pi\sqrt{d}} \quad (5)$$

## GROWTH AND MICROSCOPY DETAILS

### GaSb/AlSb/Si sample presented in Fig. 2(a&b):

The 0.3°-off and 6°-off (001) Si substrates were first prepared *ex situ* according to the procedure described in ref. [1] before being loaded into the MBE reactor. The substrate temperature was then ramped up to 800 °C at ~ 20 °C/min and then immediately cooled at the same rate down to 450 °C, without any intentional flux (all shutter cells being kept closed). MBE growth was initiated by simultaneous opening of Al and Sb shutters to grow 4 monolayers (MLs) of AlSb followed by a 5 nm thick GaSb layer. Next the temperature was ramped to 500 °C to grow a 500 nm thick GaSb layer. Three-period (5 nm GaSb/1 ML AlSb) marker superlattices were inserted after growth of 5 nm, 50 nm, 100 nm, 200 nm GaSb to track the evolution of the growth throughout the structure. These marker superlattices can be seen at large magnification on Figs. 2 a) and 2 b). The temperatures were measured by a pyrometer, and the growth rates were 0.35 ML/s for AlSb and 0.65 ML/s for GaSb.

<u>GaP/Si sample presented in Fig. 2(c&d), Fig. 3(a):</u>

GaP/Si samples presented in Fig. 2 (c&d) and Fig. 3(a) were grown by Molecular Beam Epitaxy (MBE) on a HF-chemically prepared Si(001) substrate [2]. For all the samples, the substrate has been heated at 800°C during 10 minutes to remove hydrogen at the surface, and a 10-nm thick GaP/Si deposition was performed by Migration Enhanced Epitaxy at 350°C. Subsequently, successive 4 50nm-thick GaP layers were grown by conventional MBE growth mode at increasing growth temperature: 500°C, 535°C, 565°C, and 600°C. Each MBE layer was separated by a 2-nm-thin AlGaP marker grown at the same temperature as the following GaP layer.

GaP/Si sample presented in Fig. 2(c) was grown on a nominal Si(001) substrate, with a miscut given by the manufacturer at [0±0.5°]. V/III Beam Equivalent Pressure (BEP) ratio was set to 11 during the growth.

GaP/Si sample presented in Fig. 2(d) was grown under the same condition as the sample presented in Fig. 2(c), but on a Si(001) substrate with 6° miscut towards [110] direction. V/III Beam Equivalent Pressure (BEP) ratio was set to 11 during the growth.

GaP/Si sample presented in Fig. 3(a) was grown under the same condition as the sample presented in Fig. 2(d), on a Si(001) substrate with 6° miscut towards [110] direction. But this time, V/III Beam Equivalent Pressure (BEP) ratio was set to 5.5 during the growth. Besides, after the MBE growth, an amorphous thick As capping layer was deposited on the GaP/Si(001) film at cryogenic temperature, allowing the transfer of the sample to the ultra-high vacuum STM chamber experiment, as already discussed in refs. [3,4].

<u>Transmission Electron Microscopy image of Fig. 2(a,b, c and d), S1 and S2:</u>

The GaSb/Si and GaP/Si samples have been observed in cross-sectional view by Scanning Transmission Electron Microscopy on an aberration corrected microscope Titan Themis 200. The thin foil has been prepared by FIB following the <110> zone axis (the <110> direction parallel to the surface steps linked to the 6° misorientation). The FIB preparation has been followed by a cleaning with argon milling at low voltage (1.5kV) during 9 minutes to remove the material redeposition (gallium and antimony) during the FIB process. Figures 2c, 2d, S1b, and S2 correspond to STEM Bright Field images. Figures 2a, 2b, S1.a and S1.c are Dark Field images recorded by using the (002) diffraction spot.

Histograms presented in Fig. 2 (c,d) have been drawn by counting by eyes the number, and sizes of antiphase domains observed in many TEM images. For the GaP grown on nominal (vicinal) Si substrate, TEM images examples are given in Fig. S1 (Fig. S2).

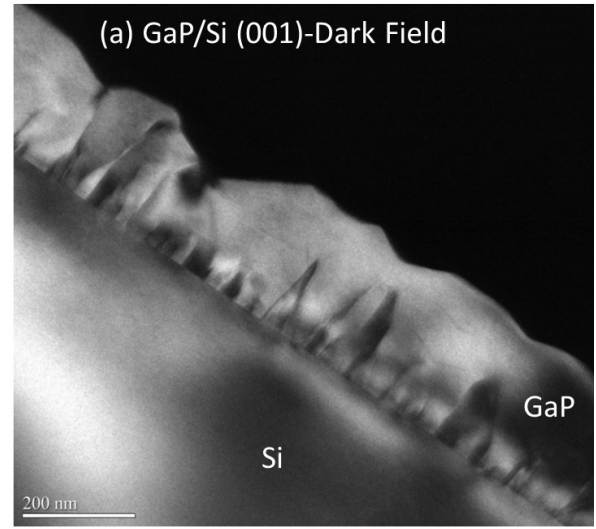

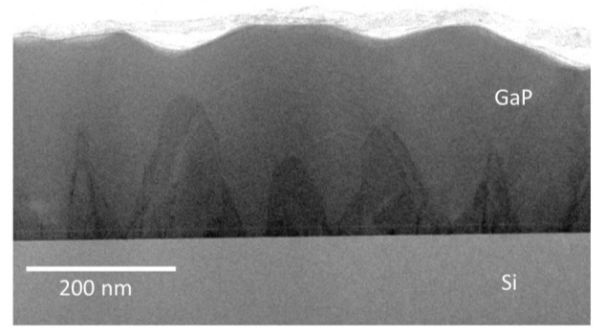

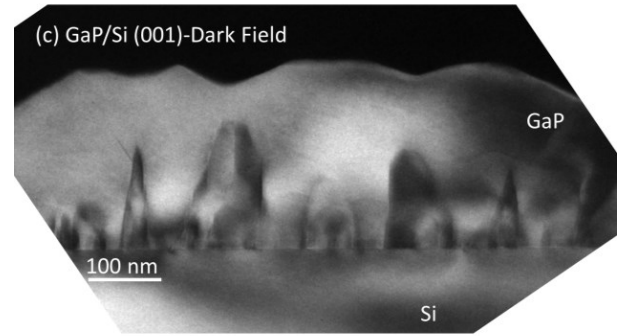

FIG. S1: Transmission Electron Microscopy images of the GaP/Si(001) sample previously described, with (a) the Dark field image of the sample over a 1μm range, (b) the bright field image corresponding to a smaller scale. (c) is the dark field image performed exactly in the same part of the sample than (b), showing the Antiphase Domains.

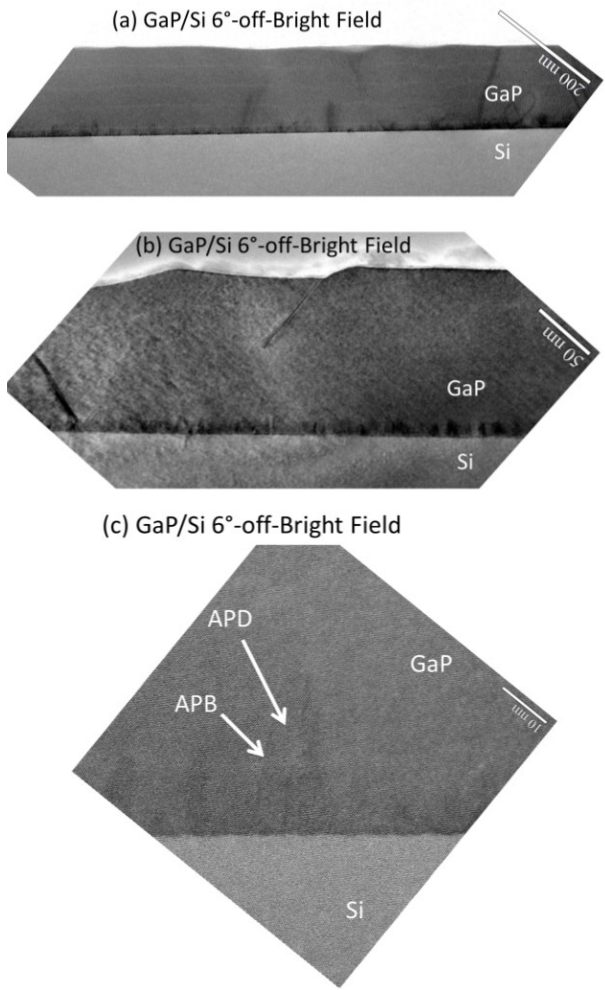

FIG. S2: Transmission Electron Microscopy images of the GaP/Si(001)-6°-off sample previously described, with the Bright-Field images at different scales (a) for a 800 nm lateral dimension, (b) with a 300 nm lateral dimension and (c) with a 60 nm lateral dimension.

Scanning Tunneling Microscopy image of Fig. 3(a):

Scanning Tunneling Microscopy (STM) was performed at room-temperature in the constant current mode of operation. Tungsten electro-chemically etched tips were used. After introduction in the ultra-high vacuum STM chamber, the protective amorphous As layer was thermally desorbed at 500°C. Raw STM images were corrected by subtraction of a basal plane.

Scanning Tunneling Microscopy image presented in Fig. 3(a) used in the manuscript to explain the antiphase domain burying is extracted from a set of experiments on different samples showing the same behavior. Fig. S3 presents a 100*100 nm² STM image obtained on the same sample.

The covering of one domain by the other is confirmed and the polarity of the two different domains can again be distinguished without any ambiguity. To check the validity and homogeneity of the process at large scale, Fig. S4(a) displays a 300*300nm² STM image of another part of the sample, typically 1 centimeter away from the observation of Fig. 3(a), showing still the same surface structure. A 150*150nm² zoom on the lower left part of Fig. S4(a) is presented in Fig.S4(b). Here extended flat (001) facets are locally observed on two neighboring domains. At this scale, the GaP(001) surface reconstruction is clearly visible and allows unambiguous identification of the local III-V [-110] direction and thus the local polarity of both domains. The GaP domain having its [-110] direction parallel to the [110] direction of the Si substrate again coalesces over the other polarity in agreement with Fig.3(a). A larger scale image on the same sample is also presented in Fig. S5 confirming the homogeneity of the process over the sample. Finally, these observations have been also made on other samples, as the GaP/Si one presented in Fig. S6, where atomic Force Microscopy on 5*5 μm² and 10*10 μm² reveals one more time the same surface morphology, corresponding to the moment where one polarity bury the other. Small inclusions of the minority polarity are still observed on these images.

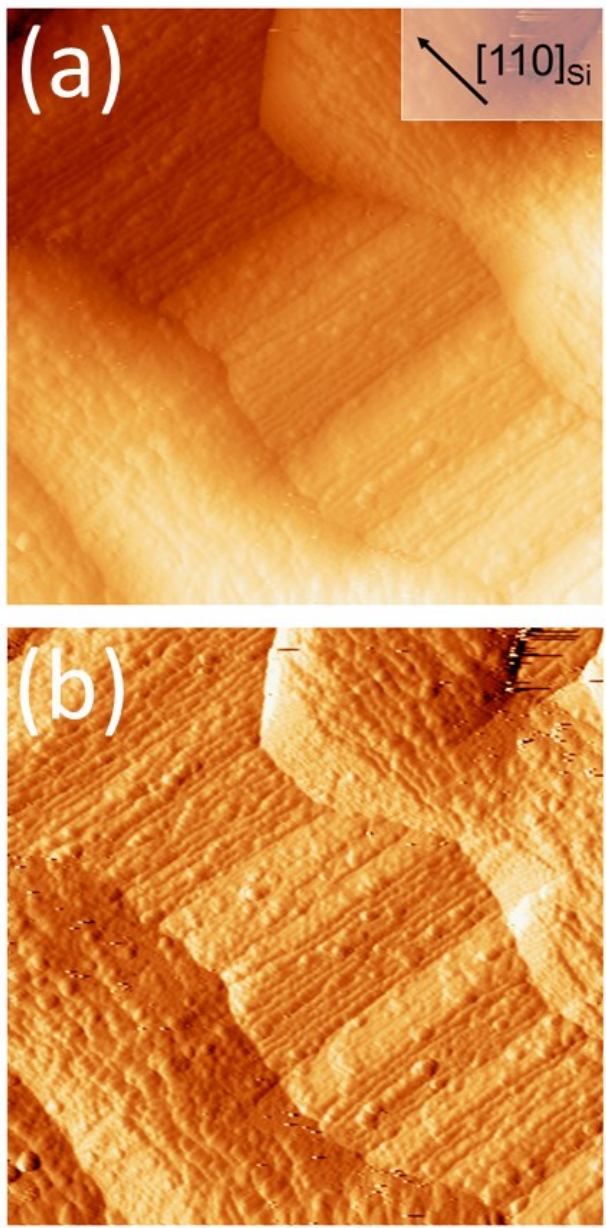

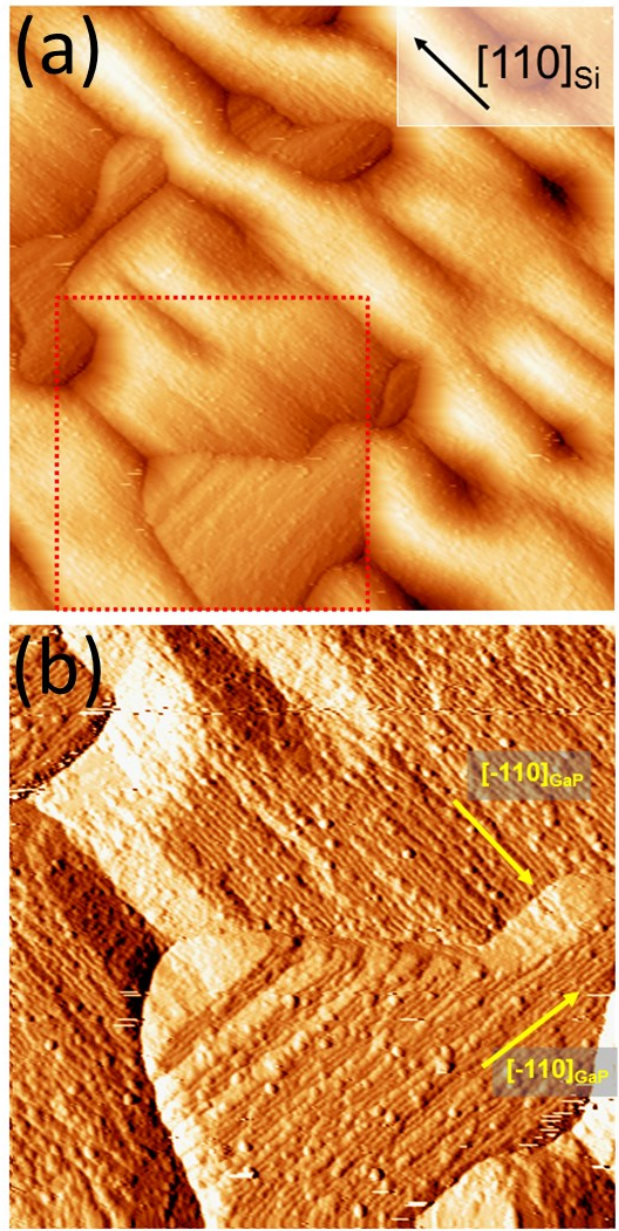

FIG. S3: 100*100 nm² Scanning Tunneling Microscopy image of the GaP/Si-6°-off sample shown in the manuscript (fig. 3(a)), (a) topography and (b) derivative of the topography along scan direction, demonstrating the burying of one domain by the other one.

FIG. S4: (a) 300*300 nm² STM image of the GaP/Si-6°-off sample shown in the manuscript (fig. 3(a)), performed elsewhere in the sample, revealing the same burying process, along the same crystal direction. Vertical color scale : 0-13.5nm. (b) 150*150 nm² zoom in the region marked in (a). The presence of (001) summital facets on the emerging antiphase domains allows unambiguous determination of the local polarity. The STM image was derived along scan direction to enhance atomic contrasts. Green arrows indicate the local [-110] GaP directions for the two antiphase domains.

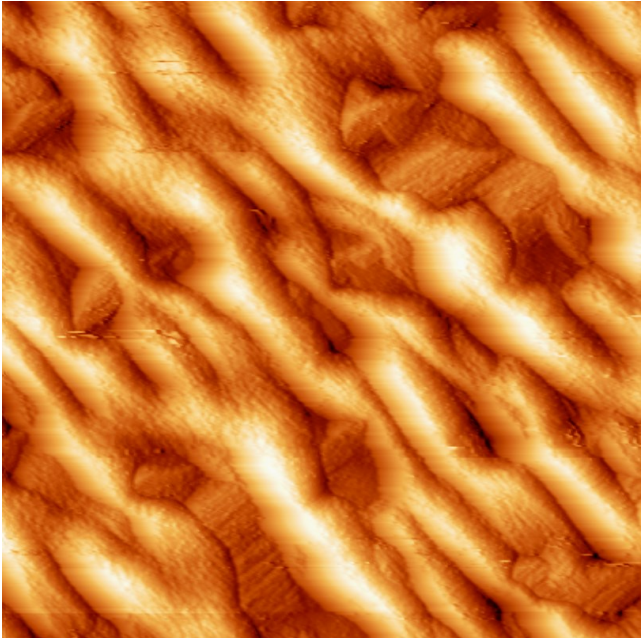

FIG. S5: 400*400 nm² Scanning Tunneling Microscopy image of the GaP/Si-6°-off sample shown in the manuscript (fig. 3(a)), revealing the same burying process at a larger scale, along the same crystal direction.

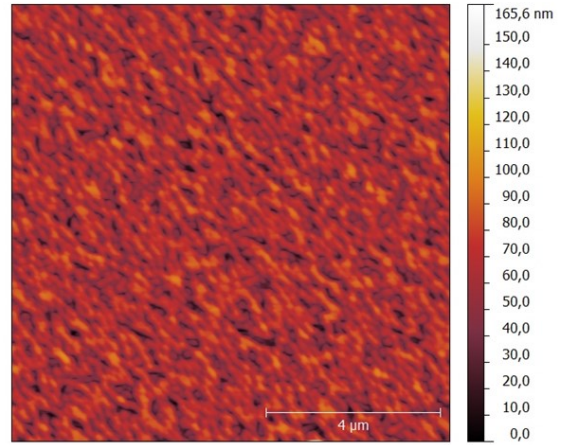

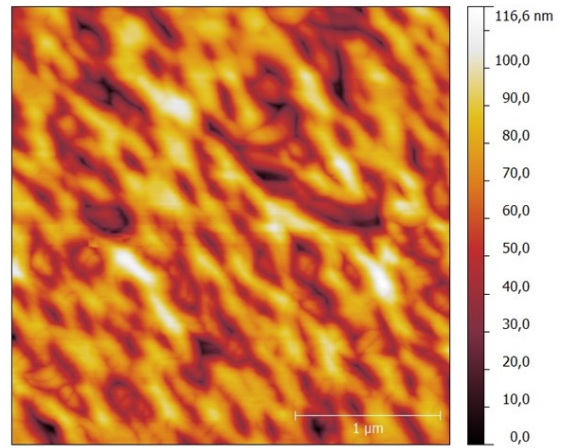

FIG. S6: Atomic Force Microscopy of another GaP/Si sample for which the burying process is ongoing. Anisotropy of the surface along the same direction can be seen on the 5*5 μm² image (a) and on the 10*10 μm² one.

## BURYING OF THE ANTI-PHASE BY THE MAIN PHASE

As mentioned in the main article, the annihilation process is not strictly speaking an annihilation, but more a burying of one phase by the other, due to different growth rates of the different phases, thanks to the miscut. As it was also reported previously that stable facets may form where the APB emerges [5], an illustration of the APB propagation and APD burying is proposed in Fig. S7. But the real situation can be quite different from this picture, depending on the different materials systems, the relative stability of high angle facets, and their growth rate.

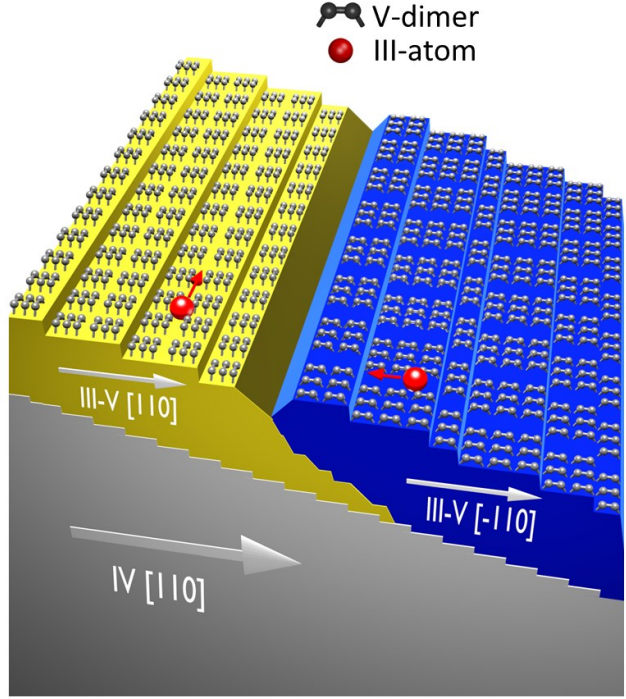

**V-dimer**
**III-atom**

III-V [110]

III-V [-110]

IV [110]

FIG. S7: Illustration of the burying process, and the propagation of the antiphase boundary when the growth rate imbalance coefficient is far from unity.

| Growth condition | Activation energy $E_a$ (eV) | | Pre-factor $R_0$ (atom.s$^{-1}$.site$^{-1}$) | |
|---|---|---|---|---|
| | A surface | B surface | A surface | B surface |
| V/III=2.0 | 4.10 | 1.96 | $8.0.10^{23}$ | $4.0.10^{11}$ |
| V/III=2.5 | 3.89 | 1.41 | $5.0.10^{22}$ | $4.7.10^{8}$ |
| V/III=4.3 | 2.63 | 1.02 | $1.5.10^{15}$ | $2.0.10^{6}$ |
| V/III=6.8 | 1.50 | 0.87 | $4.2.10^{8}$ | $3.2.10^{5}$ |

Table S1: activation energies and pre-factors extracted from the fitting of experimental data from refs. [6,7] and used to plot Fig. 4 of the manuscript.

The imbalance coefficient was then determined as a function of the temperature and V/III ratio, with the values of Table S1.

## GROWTH RATE IMBALANCE COEFFICIENT

The determination of the growth rate imbalance coefficient is conditional upon knowing experimentally the growth rates or direct step incorporation rates per site of each surfaces A and B. Experimental determination of incorporation rates was proposed in the pioneering works of Shitara et al. for MBE-grown GaAs, by using reflection high-energy electron diffraction [6,7]. In these works, they determined the direct step incorporation rate per site $R_I$, that follows an Arrhenius dependency:

$$R_I = R_0 e^{-\frac{E_a}{k_B T}} \qquad (2)$$

Therefore, the fitting of experimental data leads to the determination of a set of two parameters, that can be further extracted, namely $E_a$ the activation energy, and $R_0$ the pre-factor. The parameters used in Fig. 4 have thus been determined from the work of Shitara et al. [6,7], especially for different V/III ratio. Table S1 gives theses parameters: